\documentclass [aps,prb,preprint,superscriptaddress]{revtex4}
\usepackage{braket}
\usepackage{graphicx}
\usepackage{amsmath}
\usepackage{amsfonts}

\usepackage{color,soul} %%%%%%
\begin{document}
\title {Designing Si sphere metagratings: From perfect reflection to large-angle diffraction}

%\author[1,2]{Evangelos Almpanis}
%\author[1,2]{Emmanouil Panagiotidis}
%\author[2]{Nikolaos Stefanou}
%\author[1,*]{Nikolaos Papanikolaou}

\author{Evangelos~Almpanis} 
\affiliation{Institute of Nanoscience and Nanotechnology, NCSR
``Demokritos," \\ Patriarchou Gregoriou and Neapoleos Str., 
 Ag.~Paraskevi, GR-153~10 Athens, Greece}
 \affiliation{Section of Condensed Matter Physics, National and Kapodistrian University of Athens, Panepistimioupolis, GR-157 84 Athens, Greece}
\author{Emmanouil~Panagiotidis}
\affiliation{Institute of Nanoscience and Nanotechnology, NCSR
``Demokritos," \\ Patriarchou Gregoriou and Neapoleos Str., 
 Ag.~Paraskevi, GR-153~10 Athens, Greece}
\affiliation{Section of Condensed Matter Physics, National and Kapodistrian University of Athens, Panepistimioupolis, GR-157 84 Athens, Greece}
\author{Nikolaos~Stefanou}
 \affiliation{Section of Condensed Matter Physics, National and Kapodistrian University of Athens, Panepistimioupolis, GR-157 84 Athens, Greece}
\author{Nikolaos~Papanikolaou}\email{n.papanikolaou@inn.demokritos.gr}
\affiliation{Institute of Nanoscience and Nanotechnology, NCSR
``Demokritos," \\ Patriarchou Gregoriou and Neapoleos Str., 
 Ag.~Paraskevi, GR-153~10 Athens, Greece}

%\affil[1]{Institute of Nanoscience and Nanotechnology, NCSR “Demokritos,” Patriarchou Gergoriou and Neapoleos Str., Ag. Paraskevi, GR-153~10 Athens, Greece }
%\affil[2]{Section of Solid State Physics, National and Kapodistrian University of Athens, Panepistimiopolis Zografou, GR-157~84, Athens, Greece }
%\affil[*]{Corresponding author: n.papanikolaou@inn.demokritos.gr}

%\dates{Compiled \today}

 %           \ociscodes{(350.4238) Nanophotonics and photonic crystals; (290.4210) Multiple scattering; (230.1950) Diffraction gratings}

%%\doi{\url{http://dx.doi.org/00.0000/ao.XX.XXXXXX}}
\begin{abstract}
A thorough theoretical study of the optical properties of periodic Si nanosphere arrays is undertaken, placing particular emphasis on the synergy between electric and magnetic Mie resonances, which  occur in high-refractive-index nanoparticles and can lead to a rich variety of phenomena ranging from perfect reflection to controlled diffraction. By means of systematic calculations using the layer-multiple-scattering method that we properly extended so as to describe periodic arrays with many scatterers per unit cell, in conjunction with finite-element simulations, we optimized surfaces of Si nanospheres that efficiently channel the transmitted light into a single, first-order diffraction beam, following simple design rules based on physical insight. Our results provide compelling evidence that Huygens' metasurfaces consisting of simple Si nanosphere dimer lattices constitute a promising platform for large-angle unidirectional deflection of transmitted light
in the visible, at wavelengths shorter than the diffraction edge of the lattice.
   
\end{abstract}
%\setboolean{displaycopyright}{true}

%\begin{document}
\maketitle
\section{Introduction}

Ultrathin photonic nanostructures, known as metasurfaces, which allow the precise and efficient control of an electromagnetic (EM) wavefront, have recently been proposed and developed~\cite{flat_optics_Capasso,Review_Shalaev,Review_Kivshar,hyperbolic}. Such quasi-two-dimensional (2D) architectures can shape the amplitude, phase, polarization, or even the propagation direction of an EM wave, enabling a variety of practical utilizations such as achromatic lensing~\cite{achromatic_meta_2018,achromatic_capasso_2019}, planar holography~\cite{holog_meta_1,holog_meta_2,holog_meta_3,Wang_hologram}, information encryption~\cite{crypto_1,crypto_2}, perfect absorption~\cite{perfect_absor_1,perfect_absor_2,mimicking}, polarization switching~\cite{polariz_1,polariz_2,polariz_4}, as well as quantum~\cite{quantum} or nonlinear~\cite{nonlinear,nonlinear_2} photonics. 

One of the most intriguing properties of metasurfaces is the steering of incident light to a desired direction. A possible route in achieving beam deflection is by designing surfaces with subwavelength structuring that tailor the phase of a wavefront. The optical response of these, so-called gradient, metasurfaces is described by the generalized Snell's law~\cite{Generalized_Snell}, where a gradual shift of the transmission or reflection phase from $0$ to $2\pi$ across a particular direction results in anomalous refraction based on the Huygens-Fresnel principle~\cite{Capasso_2012,review_metas_2016,Shalaev_alldielectr_2015,Huygens_Review,Huygens_metallic,Kivshar_Huygens}. 
   However, when it comes to wide-angle deflection of the light beam, such an approach suffers from fundamental limitations which, together with fabrication difficulties due to complexity, hinder the accomplishment of simultaneous high-efficiency and large-angle light bending~\cite{Arbabi_2015,Asadchy_2016,Estakhri_2016}. 
 A different suggestion based on the equivalence principle  \cite{Huygens_metallic} was also put forward. The desired field 
 profile in space is achieved through the arrangement of electric and magnetic currents at a common interface which, for 
 visible and infrared light, can be possible by using the induced polarization currents on appropriately designed surfaces \cite{eqiv13b}. Such designs can also be realized by stacking multilayer metasurfaces with properly engineered reflection and transmission coefficients \cite{monticonnePRL}. 
An alternative route suggests the development of 2D arrangements of meta-atoms or metamolecules in a periodic lattice, taking advantage of the diffraction properties. Chiefly, such designs, often called metagratings~\cite{metagr_Alu}, have the ability to steer light into a particular diffraction channel, reaching remarkable efficiencies even at very high light-bending angles~\cite{grating_phase_diel,metagrat_metaletal,freeform,metalens,fishlike,metagrat_diel_bianisotropic}. In the transmission configuration, this property would be handy for the realization of almost perfect metalenses~\cite{metalens}. 

Obviously, metagrating designs should exhibit low Joule-heating losses together with forward scattering of an incoming EM wave. Both requirements are met when the constituent units are dielectric oligomers that fulfill Kerker's conditions for EM-wave constructive interference in the far field~\cite{Kerker,Si_dimers_1,Si_dimers_2,si_hollow_nanodisk,Si_dimer_Albella_2015,Si_dimer_Albella_2016,Si_oligomers}. 
In this respect, the optical properties of the individual oligomers should be properly combined with the characteristics of the 2D lattice to result in collective resonant modes that can funnel light through a non-zeroth order diffraction channel~\cite{freeform,metalens,fishlike,metagrat_diel_bianisotropic}.
A distinct feature of dielectric particles is that they support both electric and magnetic resonant modes, which provide unique opportunities to tailor at will the light wavefront. For example,
combinations of different optical modes occurring at the same frequency can lead to strong 
directional scattering~\cite{Kerker,Si-dimer_array,Si_dimer_Albella_2016}. 
Though such functional structures of dielectric particles have already been reported \cite{Huygens_Lattice}, the underlying design principles require a detailed understanding of the complex interaction 
mechanisms of the photonic modes, which is still missing. So far, the optimum structures were mainly obtained through 
numerical optimization algorithms, such as neural networks \cite{neural_net}, that do not enable physical insight. 

The purpose of the present work is to investigate the possibility to obtain unidirectional deflection of light into large diffraction angles with arrays of Si spheres, in transmission. Pairs of Si spheres 
have already been shown to exhibit enhanced forward scattering due to constructive interference~\cite{Kerker,Si-dimer_array,Si_dimer_Albella_2016} and arranging such sphere dimers in a lattice provides additional degrees of freedom to engineer the optical response. Here, we undertake a systematic approach to the 
design of large-angle diffraction metasurfaces. By means of rigorous, full-electrodynamic, large-scale calculations, we investigate the role of Mie resonances in high-refractive-index spherical particles and pairs of such, their interaction in a lattice, and optimize the properties of so constructed metasurfaces for unidirectional large-angle diffraction of transmitted light.
For this purpose, we extend the layer-multiple-scattering (LMS) method~\cite{lms1,lms2} to deal with different types of scatterers per unit cell, rendering it a powerful tool in the design of complex metasurfaces.

\section{Calculational Method}
The LMS method solves Maxwell's equations for layered structures of scatterers arranged with the same 2D periodicity in a homogeneous host medium, using spherical-wave expansions around each scattering centre and properly  accounting for all multiple-scattering contributions. The global scattering matrix is obtained in a plane-wave basis, through appropriate transformation from spherical to plane waves. This approach is very efficient 
for problems that involve layers of scatterers with not too large deviations from the spherical shape~\cite{gantzounis}. 
Here we shall briefly describe an extension of the method to treat layers with many scatterers per unit cell.

In an isotropic and homogeneous medium with a relative electric permitivity $\epsilon$ and magnetic permeability $\mu$, the solutions of the electrodynamic equations at given angular  frequency $\omega$ are transverse plane EM waves propagating with wave vector ${\bf q}$ ($q= \omega \sqrt{\epsilon \mu} /c$, $c$ being the speed of light in vacuum), which have an electric-field component

\begin{equation}\label{eq:plw}
\mathbf{E} _{\mathbf{q} p} (\mathbf{r}) = \mathrm{Re} \big[ \mathbf{E}_{0} (\mathbf{q})
\exp [\mathit i (\mathbf{q} \cdot \mathbf{r}-\omega t)] \big],
\end{equation}
with 
$\mathbf{E}_{0} (\mathbf{q}) = E_0 \mathbf{e} _{p} (\mathbf{q})$, 
where $\mathbf{e}_{p} (\mathbf{q})$,
$p=1,2$ denotes the polar and azimuthal unit vectors associated with $\mathbf{q}$
and defines two independent linear polarization states.

To solve the problem of EM scattering by a single particle, we expand the incident wave into regular at the origin transverse spherical vector wave functions

\begin{equation}\label{eq:vecsph1}
\mathbf{J} _{H \ell m } (\mathbf{r}) = j_{\ell} (qr) \mathbf{X} _{\ell
m} (\mathbf{r}) 
\end{equation}

and 
\begin{equation}
\mathbf{J}_{E \ell m}({\mathbf r})=\frac{\mathit i}{q}
\nabla \times j_{\ell} (qr) \mathbf{X}_{\ell m} (\mathbf{r})
\label{eq:vecsph3}
\end{equation}
at the given frequency, where $j_{\ell}$ are spherical Bessel functions and ${\mathbf X}_{\ell m}$ vector spherical harmonics defined as
\begin{equation}
\sqrt{\ell(\ell +1)} {\mathbf X}_{\ell m}({\mathbf r})=
-i{\mathbf r} \times \nabla Y_{\ell m}({\mathbf r})
  \label{eq:Xlm}  
\end{equation}
with $Y_{\ell m}({\mathbf r})$ being the usual spherical harmonics.
We similarly define spherical vector functions $\mathbf{H} _{H \ell m}$ and
$\mathbf{H} _{E \ell m}$, by replacing in the above Eqs. (\ref{eq:vecsph1}) and (\ref{eq:vecsph3}) the spherical Bessel functions $j_{\ell}$ by spherical Hankel functions $h ^{+}_{\ell}$, which are appropriate for outgoing waves. In what follows, we shall employ an index ${\mathrm L}$ to denote, collectively, the indices $P \ell m$, where $P=E,H$, $\ell \ge 1$, and $m=-\ell, -\ell +1, \ldots \ell$.

In the presence of a scattering particle, the total field is composed of the  
incoming wave, $\mathbf{J}_{\mathrm{L}}(\mathbf{r})$, plus the scattered wave, 
$\mathbf{H}_{\mathrm{L}}(\mathbf{r})$, with
appropriate coefficients
\begin{equation}
\mathbf{E}(\mathbf{r})=\sum_{\mathrm{L}} \left[a_{\mathrm{L}}^{0}
\mathbf{J}_{\mathrm{L}}(\mathbf{r}) + a_{\mathrm{L}}^{+}
\mathbf{H}_{\mathrm{L}}(\mathbf{r}) \right] \;. \label{eq:genfield}
\end{equation}
The expansion coefficients $a_{\mathrm{L}}^{0}$ and $a_{\mathrm{L}}^{+}$ are related to each other through the 
scattering matrix $\mathbf t$ of the individual scatterer. For isotropic spherical particles, this matrix is diagonal in ${\mathrm L}$, with elements $t_{\mathrm L}$ which are independent of $m$ and can be evaluated analytically. In this case we have $a_{\mathrm L}^{+} = t_{\rm L} a_{\mathrm L}^{0}$.

We now consider a 2D lattice with many particles per unit cell at positions 
${\mathbf R}_{n \beta}= {\mathbf R}_n+ {\mathbf r}_{\beta}$,
where ${\mathbf R}_n=n_1 {\mathbf a_1} + n_2 {\mathbf a_2}$ with $n_1, n_2$ integer numbers are the Bravais lattice translations and ${\mathbf r}_{\beta}$ are appropriate non-primitive vectors within the unit cell.
Using the addition theorem for spherical harmonics and Bessel functions, an outgoing spherical wave about a given scatterer can be expanded into incoming spherical waves around a different center through appropriate propagator functions
$\Omega$ at the given frequency, as follows
\begin{equation}
\mathbf{H}_{\mathrm L'} (\textbf{r}-\textbf{R}_{n'\beta'}) =\sum_{\mathrm L} {\Omega}_{\mathrm{ LL'}}^{n\beta,n'\beta'}\mathbf{J}_{\rm L}({\textbf{r}-\textbf{R}_{n\beta})}.
\label{eq:addtheo}
\end{equation}
Explicit expressions for the propagator functions $\Omega$ can be found elsewhere \cite{lms1,LmsMacLaren}.
Scattering from this array of scatterers, after summing up all multiple-scattering events, can be described through a collective $T$ matrix given by
\begin{equation}
{\mathbf T} = ({\mathbf t}^{-1}- {\mathbf \Omega})^{-1},
\label{eq:T}
\end{equation}
where the $t$ matrix of individual spherical scatterers is diagonal with matrix elements $\delta_{\beta \beta'} t_{\rm L}^{\beta }$
and
\begin{equation}
{\Omega}_{\beta {\rm L},{\beta}'{\rm L}'} (\textbf{k}_{\parallel}) =\sum_{n'} { {\Omega}_{\rm{LL'}}^{n\beta,n'\beta'} \exp [-{\mathit i} {\textbf k_{\parallel}} \cdot (\textbf{R}_{n}-\textbf{R}_{n'})]}
\label{eq:addtheo1}
\end{equation}
is the 2D Fourier transform of the propagator functions, with $\mathbf{k}_{\parallel}$ being the in-plane component of the wave vector $\mathbf{q}$ reduced within the surface Brillouin zone of the given 2D lattice. The infinite sum in Eq.~(\ref{eq:addtheo1}) is efficiently calculated using Kambe's summation method \cite{LmsMacLaren}. For a basis of $N_{b}$ particles per unit cell and angular momentum cut-off $\ell_{\mathrm{max}}$, the matrices in Eq.~(\ref{eq:T}) have dimensions 
$2\ell_{\mathrm{max}}(\ell_{\mathrm{max}}+2)N_{b} \times 2\ell_{\mathrm{max}}(\ell_{\mathrm{max}}+2)N_{b}$. 
In the case of submicron spherical particles, the angular momentum expansions involved converge fast. For example, in the calculations carried out in the present work, $\ell_{\mathrm{max}}=5$
is sufficient to obtain excellent convergence in all cases studied. 

Let the plane wave, described by Eq.~(\ref{eq:plw}), be incident on the array of spheres. We can always write the in-plane component of its wave vector as $\mathbf{q}_{\parallel}= \mathbf{k}_{\parallel}+ \mathbf{g}'$, where the reduced wave vector $\mathbf{k}_{\parallel}$ lies in the surface Brillouin zone and remains invariant in the (multiple) scattering process due to the periodicity of the structure, and $\mathbf{g}'$ is a certain reciprocal-lattice vector. 
In turn, the scattered field at the given frequency has the form of plane waves propagating or decaying in the positive ($+$) and negative ($-$) $z$ direction, with wave vectors
\begin{equation}
{\mathbf K}_{\mathbf g}^{\pm}=\big( \mathbf k_{\parallel} + {\mathbf g} , \pm[q^2-({\mathbf k_{\parallel}+{\mathbf g}})^2]^{1/2}   \big)
\label{eq:Kg},
\end{equation}
where $\mathbf{g}$ are vectors of the reciprocal lattice, which correspond to the different diffracted beams.

The transmission and 
reflection coefficients of the array of scatterers, in the specific plane-wave basis, are obtained using the collective $T$ matrix defined in Eq.~(\ref{eq:T}) by~\cite{lms1,lms2,LmsMacLaren}
\begin{equation}
S_{sp\mathbf{g},s'p'\mathbf{g}'}=\delta_{ss'} \delta_{pp'} \delta_{\mathbf{g}\mathbf{g}'}+ \sum_{\beta \mathrm{L}} \sum_{\beta' \mathrm{L}'} \Delta_{sp\mathbf{g}}^{\beta \mathrm{L}} T_{\beta \mathrm{L}, \beta' \mathrm{L}'} A_{\beta' \mathrm{L}'}^{s'p'\mathbf{g}'} \;,
\label{eq:Ms}
\end{equation}
where $s,s'=\pm$ and 

\begin{equation}
A^{sp\mathbf{g}}_{\beta H \ell m} = 
 4\pi i^{\ell}(-1)^{m+1} \exp(\textit i\mathbf K_{\mathbf g}^s \cdot {\mathbf r}_{\beta})
 \mathbf{X} _{\ell -m} (\mathbf{K}_{\mathbf{g}}^s)\cdot \mathbf{e}_{p} (\mathbf{K}_{\mathbf{g}}^{s}) 
 \end{equation}

\begin{equation}
A^{s p {\mathbf g}}_{\beta E \ell m} =
 4\pi {\mathit i}^{\ell} (-1)^{m+1}
 \exp({\textit i} {\mathbf K}_{\mathbf g}^s \cdot {\mathbf r}_{\beta})
 [\mathbf{X} _{\ell -m} (\mathbf{K} _{\mathbf{g}} ^s) \times
\mathbf{K} _{\mathbf{g}} ^s]\cdot \mathbf{e}_{p} (\mathbf{K}_{\mathbf{g}} ^{s}) 
\end{equation}

\begin{equation}
\Delta^{\beta H \ell m}_{sp {\mathbf g}} 
	= \frac{2 \pi (-{\mathit i}) ^{\ell}} 
	{q A_0 K _{{\mathbf g}; z}^{+}}
	\exp(-i\mathbf K_{\mathbf g}^s \cdot {\mathbf r}_{\beta})
	\mathbf{X} _{\ell m} ({\mathbf{K}_\mathbf{g}} ^s)
	\cdot \mathbf{e}_{p} (\mathbf{K}_{\mathbf{g}} ^{s})
	\end{equation}
	
\begin{equation}
\Delta^{\beta E \ell m}_{sp {\mathbf g}} = \frac{2 \pi 
(-{\mathit i})^{\ell}} {q^2 A_0 K _{\mathbf{g}; z}^{+}}
\exp(-i\mathbf K_{\mathbf g}^s \cdot {\mathbf r}_{\beta}) 
[\mathbf{X} _{\ell m} (\mathbf{K}_\mathbf{g}^s) \times
 \mathbf{K}_\mathbf{g}^s] \cdot \mathbf{e}_{p} ({ \mathbf K}_{ \mathbf g}^{s})
\end{equation}
with $A_0$ being the surface of the unit cell.

For a plane wave incident on the array of scatterers, say, along the positive $z$ direction with wave vector $\mathbf{K}_{\mathbf{g'}}^{+}$ and polarization $p'$, the total transmittance and reflectance are given by

\begin{equation}
\mathcal{T} = \sum_{p \mathbf{g}} |S_{+p\mathbf{g},+p'\mathbf{g}'}|^{2} \frac{K_{\mathbf{g}z}^{+}}{K_{\mathbf{g}'z}^{+}}
\end{equation}
and
\begin{equation}
\mathcal{R} = \sum_{p \mathbf{g}} |S_{-p\mathbf{g},+p'\mathbf{g}'}|^{2} \frac{K_{\mathbf{g}z}^{+}}{K_{\mathbf{g}'z}^{+}}\;,
\end{equation}
respectively, while the absorbance is calculated by requiring energy conservation, $\mathcal{A} = 1 - \mathcal{T} - \mathcal{R}$. We note in passing that it is straightforward to generalize the LMS method to deal with multiple layers of 2D arrays with many scatterers per unit cell, provided that they have the same
periodicity~\cite{lms1,lms2}.
This method is accurate, rapidly converging, and ideally suited for arrays of spheres. One of its considerable advantages is that it allows
to directly access the character of the optical modes of the individual particles (dipolar, quadrupolar, octapolar, etc.) and gain physical insight on the formation of collective modes when these particles are arranged in a lattice.

\begin{figure}[htbp]
	\centering
	\fbox{\includegraphics[width=0.5\linewidth]{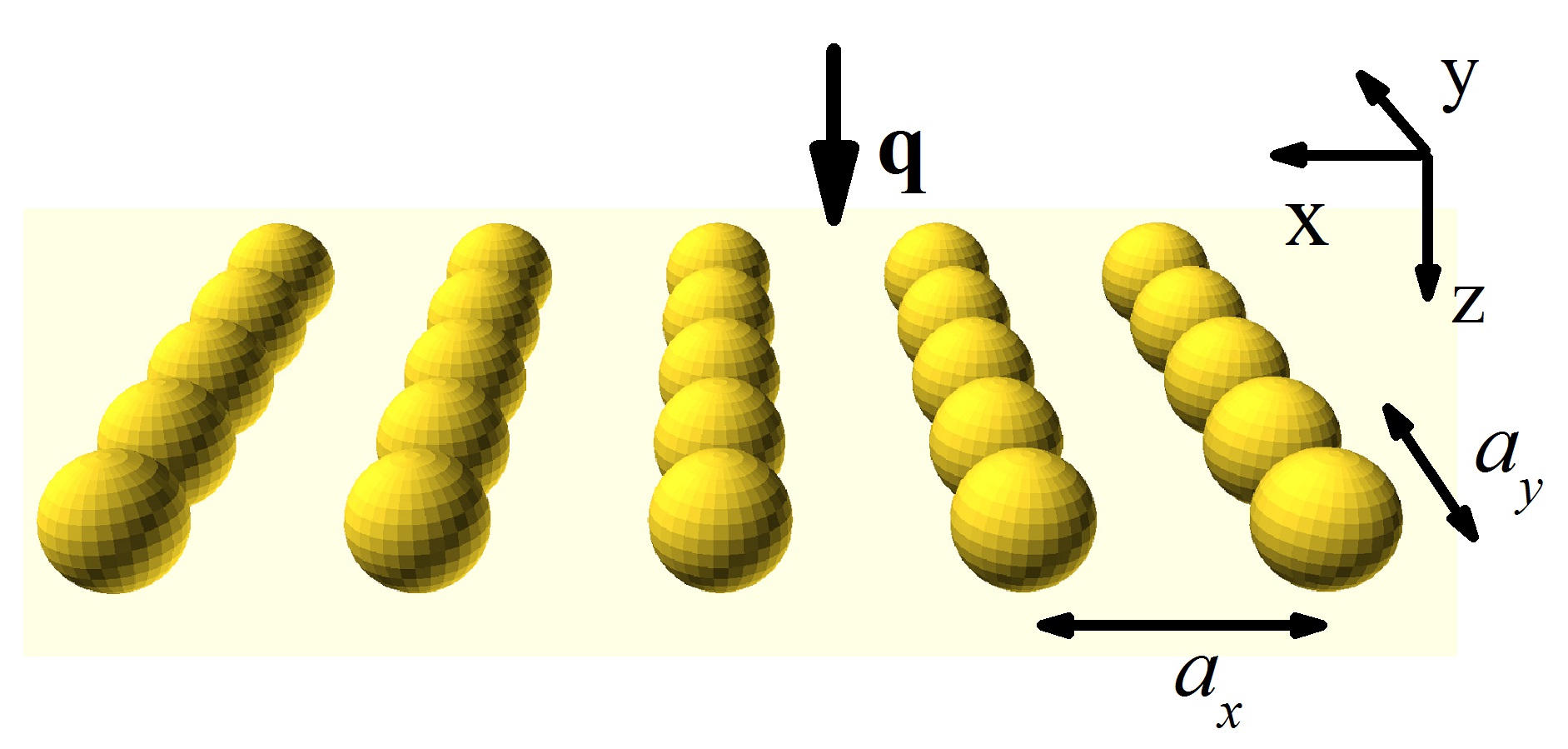}}
	\caption{Schematic of an array of spheres in a rectangular lattice with lattice constants $a_x$, $a_y$ in the $x$ and $y$ direction, respectively. 
	}
	\label{fig_lattice}
\end{figure}

\section{Results and discussion}

In dielectric particles, the existence of electric- as well as magnetic-type Mie resonances have attracted considerable attention \cite{Kerker,GeNatCom12,FuNatCom13,TrSciRep15} because their proper combination can lead to constructive or destructive interference and directional scattering~\cite{MiMiNS15, Si_dimers_2, Si-dimer_array}. In arrays of such particles, multiple-scattering effects may also be significant and are sometimes  discussed in terms of interaction between Mie modes \cite{diel_nongradient}. For example, a square array of Si spheres can be designed as a total reflector  at wavelengths close to the Mie resonances \cite{Sireflector}. In many studies of periodic arrays, the discussion is focused on wavelengths above the lattice diffraction edge, where only a single specular beam exists 
and the optical response can be related to the complex transmission and reflection coefficients of zeroth order, or to the impedance and effective optical constants~\cite{Review_Kivshar}.
At wavelengths below the diffraction edge,  
optical power is generally distributed in the different diffraction orders. By manipulating the hybridization between Mie modes, it is possible to control the intensity of light in each diffracted beam. There are several proposals for reflection metagratings \cite{reflmeta1,reflmeta4} and recent efforts to develop design rules for such architectures came to the conclusion that anisotropic unit cells supporting at least four resonances are required to achieve control of the wavefront and high efficiency \cite{metagrat_diel_bianisotropic}. For transmission gratings, \cite{trans1,trans3,monticonnePRL} one
approach is to block the specular transmission and employ asymmetric unit cells to funnel all optical power in a given diffraction order. 
In the following, we will discuss an implementation of this strategy using lattices of Si spherical particles.

\subsection{Mie modes and diffraction}

\begin{figure}[htbp]
	\centering
	\fbox{\includegraphics[width=0.5\linewidth]{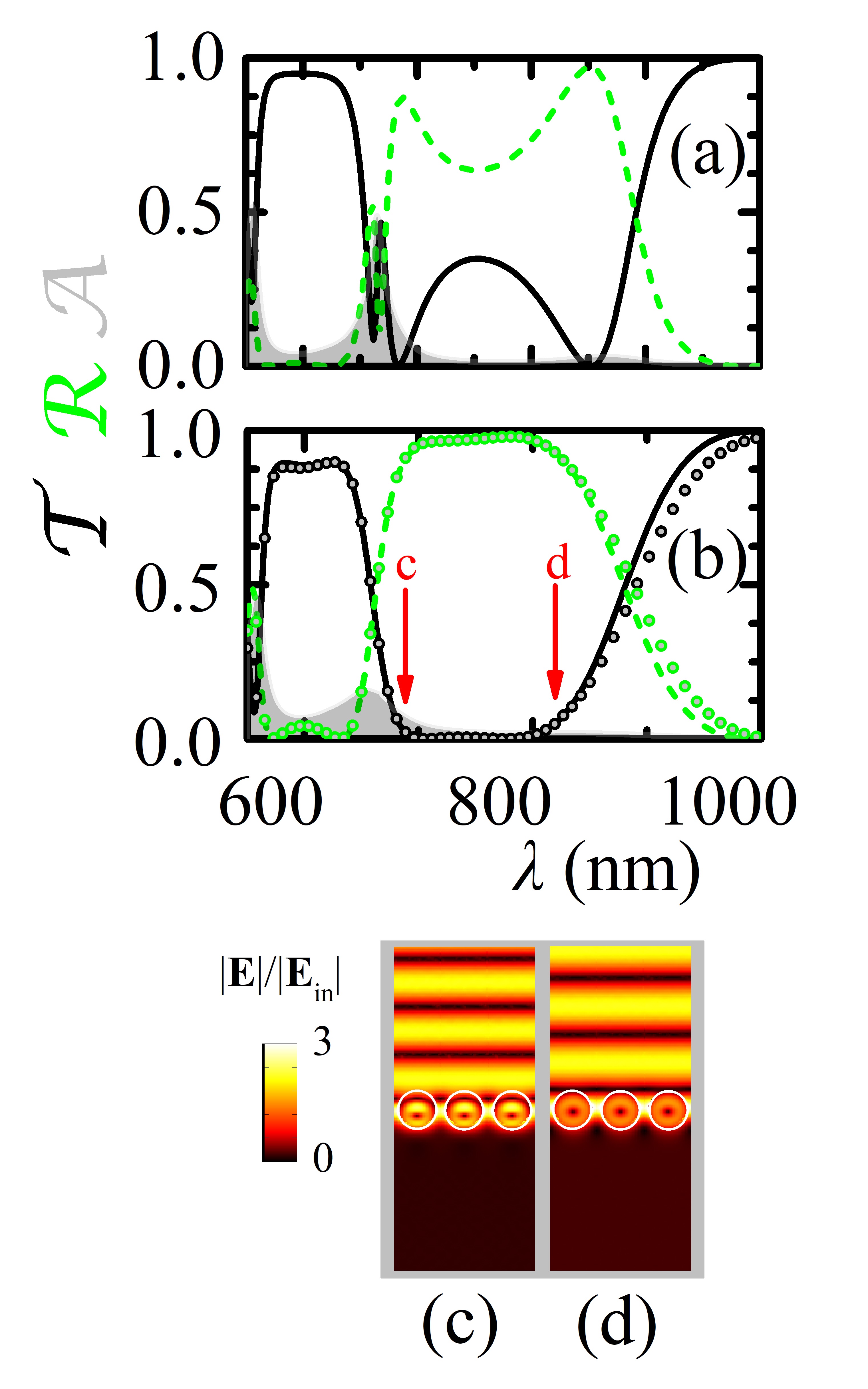}}
	\caption{(a): Optical spectra of a square array (latice constant $a_{x,y}=400$~nm) of
	Si spheres (radius $r_1=$120~nm) in air, illuminated by normally incident light (see Fig. \ref{fig_lattice}). The transmittance, $\mathcal{T}$, reflectance, $\mathcal{R}$, and absorbance, $\mathcal{A}$, are displayed by solid black lines, dashed green lines, and gray-shaded areas, respectively.
(b): The same as in (a), for $a_{x,y}=300$~nm. The circles show the results of finite-element calculations using the COMSOL Multiphysics software package. 
The normalized field profiles at the wavelengths indicated by the arrows in (b) are shown in (c) and (d).
	}
	\label{figure002}
\end{figure}

First we consider square arrays of Si spheres of the same radius, $r_1=$120~nm, in air, as schematically
shown in Fig.~\ref{fig_lattice}. In our calculations we use optical constants fitted to 
the measured data for crystalline Si~\cite{AspnesSi}. 
The transmission, reflection and absorption spectra for a lattice constant $a_{x,y}=400$~nm are shown in Fig.~\ref{figure002}(a). The displayed spectra are dominated by two pronounced reflection peaks, the dipolar magnetic (DM) close to $850$~nm and the dipolar electric (DE) about $700$~nm, which  are slightly blue-shifted from the positions of the corresponding Mie resonances of the isolated sphere \cite{Si_sphere_Mie}.
By reducing the lattice constant to $a_{x,y}=300$~nm, these peaks broaden due to the increased interparticle coupling and merge into each other forming a broad, strongly reflecting band ranging from $700$ to $800$~nm, as can be seen in Fig.~\ref{figure002}(b). The phase acquired by the electric field upon reflection, inside the reflection band, varies with the wavelength according to the character of the Mie resonances. This is clearly visible in the electric field profiles at the two edges of this reflection band, depicted in Figs.~\ref{figure002}(c),~(d).  
It is worth-noting that the spectra calculated by the LMS method [lines in Fig.~\ref{figure002}(b)] agree very well with the results obtained at increased computational cost using the the finite-element method as implemented in the COMSOL Multiphysics software package [circles in Fig.~\ref{figure002}(b)] .

The role of the lattice in the formation of a reflection band can be further elucidated by arranging the Si spheres in a rectangular array with $a_x=700$~nm, and $a_y=300$~nm (see Fig.~\ref{fig_lattice}), where at normal incidence we expect the emergence of diffracted beams in the $x$ direction for $\lambda < 700$~nm.
In general, the diffraction angle for a particular diffraction order $m_i$ along the direction with period $a_i, (i=x,y)$ at wavelength $\lambda$ is given by 
\begin{equation}
\varphi_{m_i} = \arcsin({m_i\lambda}/a_i n)
\label{eq:diff}
\end{equation}
with $n$ being the refractive index of the embedding medium.

\begin{figure}
	\centering
	\fbox{\includegraphics[width=8cm]{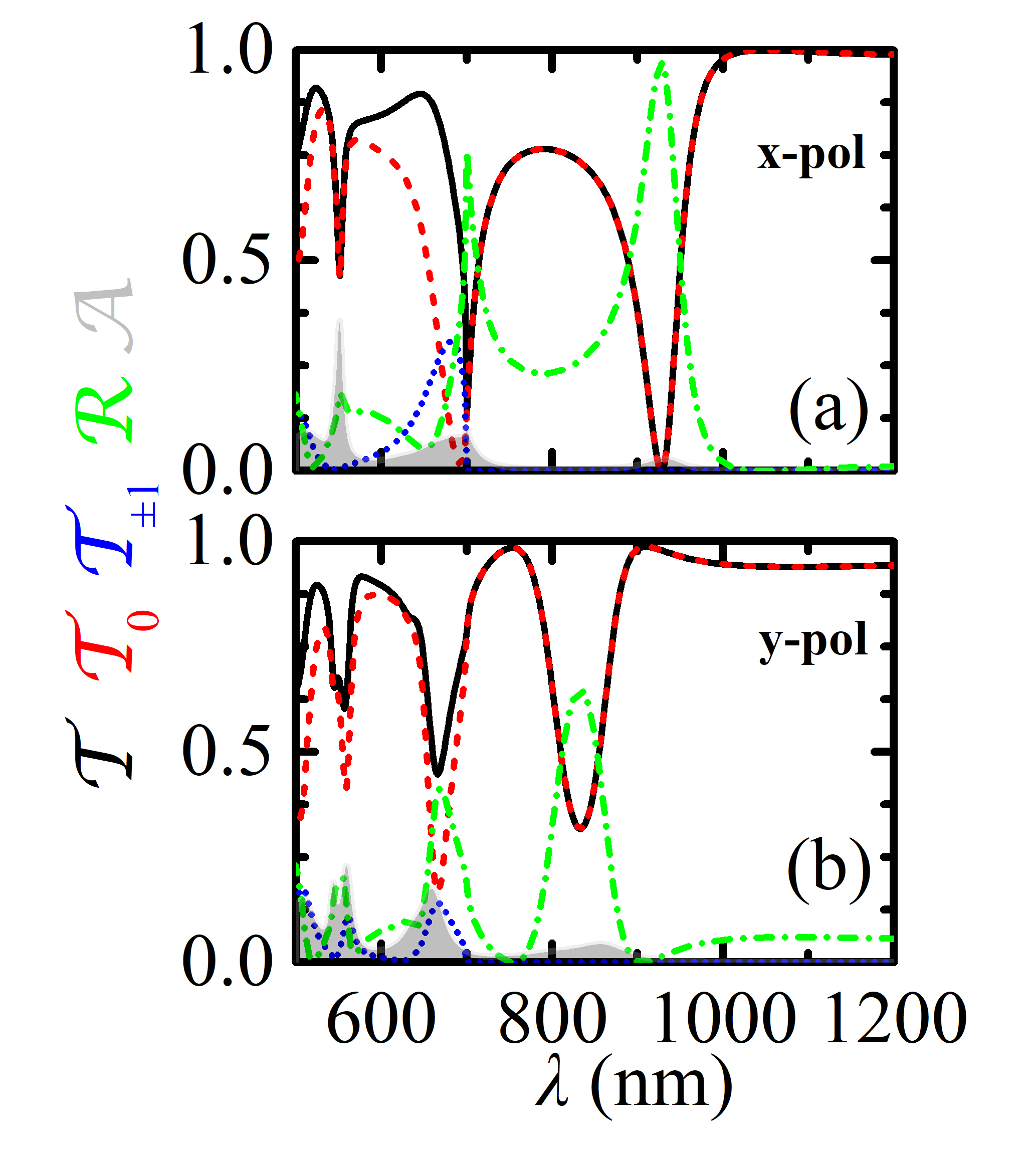}}
	\caption{Reflectance, $\cal R$ (dash-dotted green lines), absorbance, $\cal A$ (gray-shaded areas), total transmittance, $\cal T$ (solid black lines), specular transmittance, ${\cal T}_0$ (dashed red lines),
	and first-order diffracted transmitance along the $x$ direction, ${\cal T}_{\pm 1}$ (dotted blue lines) 
	of an array of Si spheres of radius $r_1=120$~nm, arranged on a rectangular lattice with $a_x=700$~nm and $a_y=300$~nm (see Fig. \ref{fig_lattice}). Light is incident normally with (a): $x$ polarization and (b): $y$ polarization. }
	\label{figure003}
\end{figure}

The reflection spectra for normally incident light, polarized along the long-period axis $x$ (x-pol) and along the short-period axis $y$ (y-pol),
are shown in Fig.~\ref{figure003}(a) and (b), respectively. 
For $x$-polarized light, the DM Mie mode is responsible  for the strong reflection  peak at longer wavelengths, close to $\lambda= 920$~nm (dash-dotted green lines), while a sharper and smaller peak associated with the DE mode appears very close to the diffraction edge at $700$~nm. The Mie modes can be traced clearly in the specular transmission component, ${\cal T}_0$, which is useful for wavelengths below the diffraction edge where the intensity of light distributed in the diffraction channels in the $x$ direction, ${\cal T}_{\pm 1}$, becomes significant and the total transmission and reflection spectra are harder to interpret. 
 For $y$-polarized light, the Mie modes give weaker reflection peaks. The DM Mie resonance peak is found to be blue-shifted compared to the $x$-polarized spectra, while the DE modes cause a significant drop in ${\cal T}_0$, similar to the spectra for the $x$ polarization.    
 
\begin{figure}
  \centering	
  	\fbox{\includegraphics[width=0.5\linewidth]{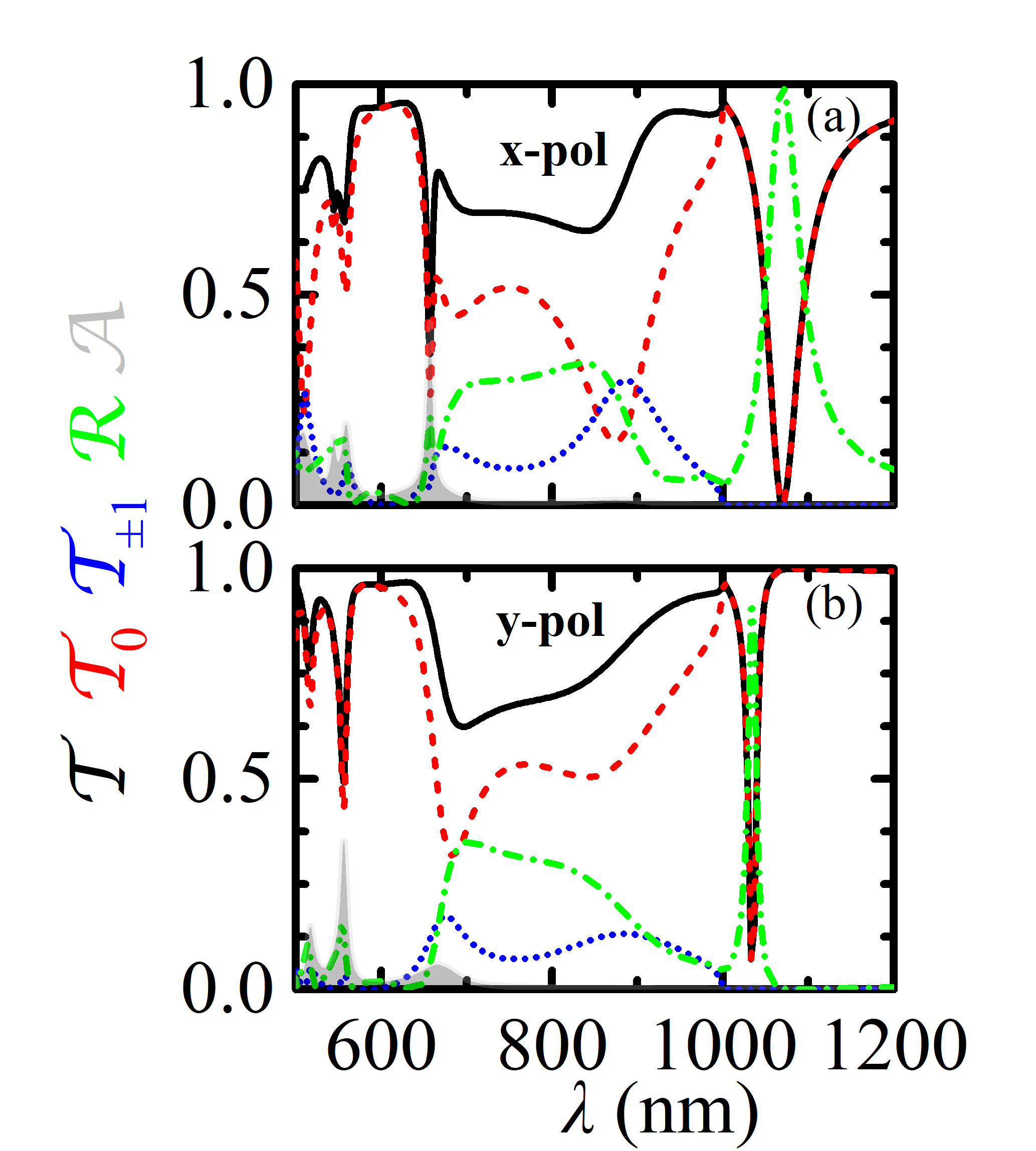}}
  	\caption{The same as in Fig.~\ref{figure003}, for $a_x=1000$~nm and $a_y=300$~nm (see Fig. \ref{fig_lattice}). }
  	\label{figure004}
  \end{figure}

By further increasing the long lattice constant, $a_x$, the DM Mie modes also move below the diffraction edge. 
Relevant optical spectra for Si spheres 
of the same radius, $r_1=120$~nm, arranged on a lattice with 
$a_x=1000$~nm and $a_y=300$~nm, are depicted in Fig.~\ref{figure004} for both polarizations.
For $x$-polarization [see Fig.~\ref{figure004}~(a)], the specular transmission,  ${\cal{T}}_0$, is suppressed at the positions of the Mie resonances, close to 880~nm for the DM mode and 680~nm for the DE mode, with the effect being stronger for the DM mode, while the first-order diffracted light along $x$, ${\cal T}_{\pm 1}$, is enhanced at these frequencies. The spectra exhibit similar features also for $y$-polarized 
light, the only difference being that the ${\cal T}_0$ minima are now close to the DE resonance. Due to the symmetry of the structure, diffraction is equally distributed in the $m_x=\pm 1$ channels.
We note that reflection is not very high in the spectral region between the DM and DE Mie resonances, but  
a narrow total reflection band
is present above the diffraction edge, close to $\lambda=1050$~nm. This is ascribed to a mode strongly localized in the sphere rows, which has the characteristics of a leaky slab mode that couples to the continuum and can also lead to a strong reflection band
\cite{Leaky_mode}. These latter modes, which appear only above the diffraction edge, are not related to a particular Mie resonance
 and can be shifted to longer wavelengths simply by increasing $a_x$.

To summarize, we have shown that, in simple 2D arrays of Si spheres, the presence of Mie resonances can lead to destructive interference in the specular transmission channel. Moreover, at wavelengths where diffraction takes place, the layers can be highly transmissive at the Mie resonances but light escapes through the higher-order diffraction channels. 

\subsection{A Si sphere dimer}

The scattering properties of dielectric dimers and multiparticle clusters with overlapping electric and magnetic Mie resonances (Kerker's condition) have drawn significant attention and several realizations were considered, since constructive interference results in enhanced forward scattering~\cite{Kerker}.

\begin{figure}[htbp]
\centering
\fbox{\includegraphics[width=0.5\linewidth]{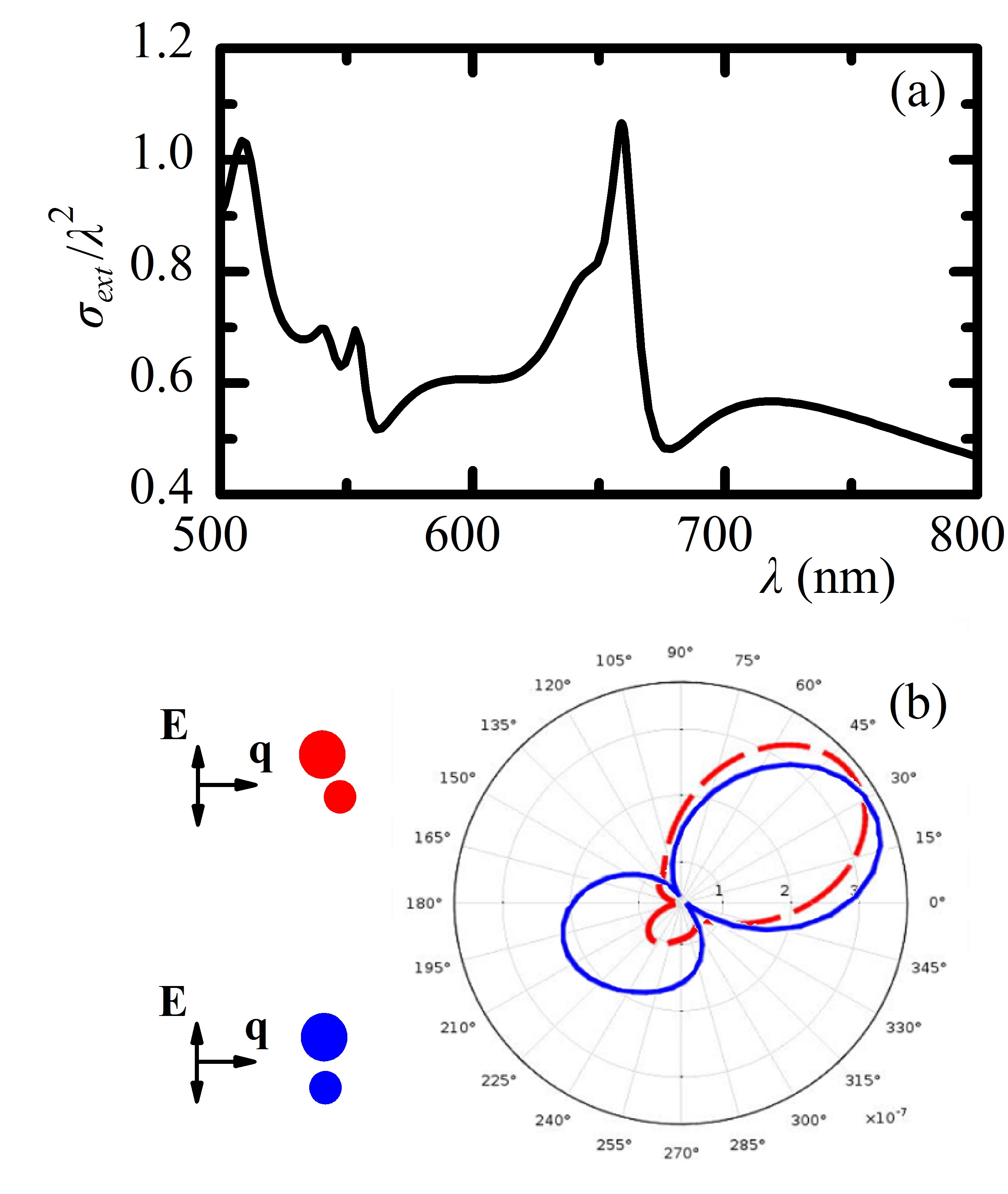}}
\caption{(a): Normalized extinction cross section, $\sigma_{\mathrm{ext}}$, of a Si 
sphere dimer with radii $r_1=120$~nm and $r_2=80$~nm, separated by $10$~nm, for light incident 
normally to (and polarized along) the dimer axis. (b): Far-field plots at $\lambda=650$~nm,
for the geometries shown on the left: Light incident normally and polarized along the dimer axis (solid blue line); light incident on a dimer tilted by $28^o$ around the 
center of the larger sphere, 
where forward scattering and deflection angle are maximized (dashed red line).
}
\label{figure005}
\end{figure}

The scattering cross section of an isolated pair of Si spheres \cite{Si-dimer_array,Si_dimer_Albella_2016,Si_dimers_2} with radii $r_1=120$ nm and $r_2=80$~nm, separated by a 10~nm air gap, calculated using the COMSOL Multiphysics software package, is shown in Fig.~\ref{figure005}(a).
  The radii were chosen so that the DM mode of the
  smaller sphere is in the same spectral region as the
  DE mode of the
  larger sphere. 
  Mode overlap leads to enhanced forward scattering 
  at wavelengths around $650$~nm as shown in the far field plots [Fig.~\ref{figure005}(b)], while the asymmetry of the dimer deflects light towards the
  direction of the larger sphere. The structure can be
  optimized to reduce backscattering by tilting the dimer  
  by an angle $\theta=28^o$ around the center of the larger 
  sphere. As shown with the dashed red lines in Fig.~\ref{figure005}(b), this tilted configuration leads to stronger forward scattering and larger deflection angle while the scattering cross section spectrum changes only slightly (not shown). This suppression of 
  backscattering is accompanied by hot-spots of the electric field in the region between the spheres \cite{MiMiNS15}.

\subsection{Surfaces for efficient large-angle deflection of light}
 
Dielectric oligomers with controlled, directional, scattering as the simple Si sphere dimer considered here are interesting building blocks of periodic metagratings, which efficiently deflect light towards the direction of a particular
diffraction order~\cite{achromatic_meta_2018,metagrat_metaletal,metagrat_diel_bianisotropic}.

\begin{figure}[htbp]
\centering
\fbox{\includegraphics[width=0.5\linewidth]{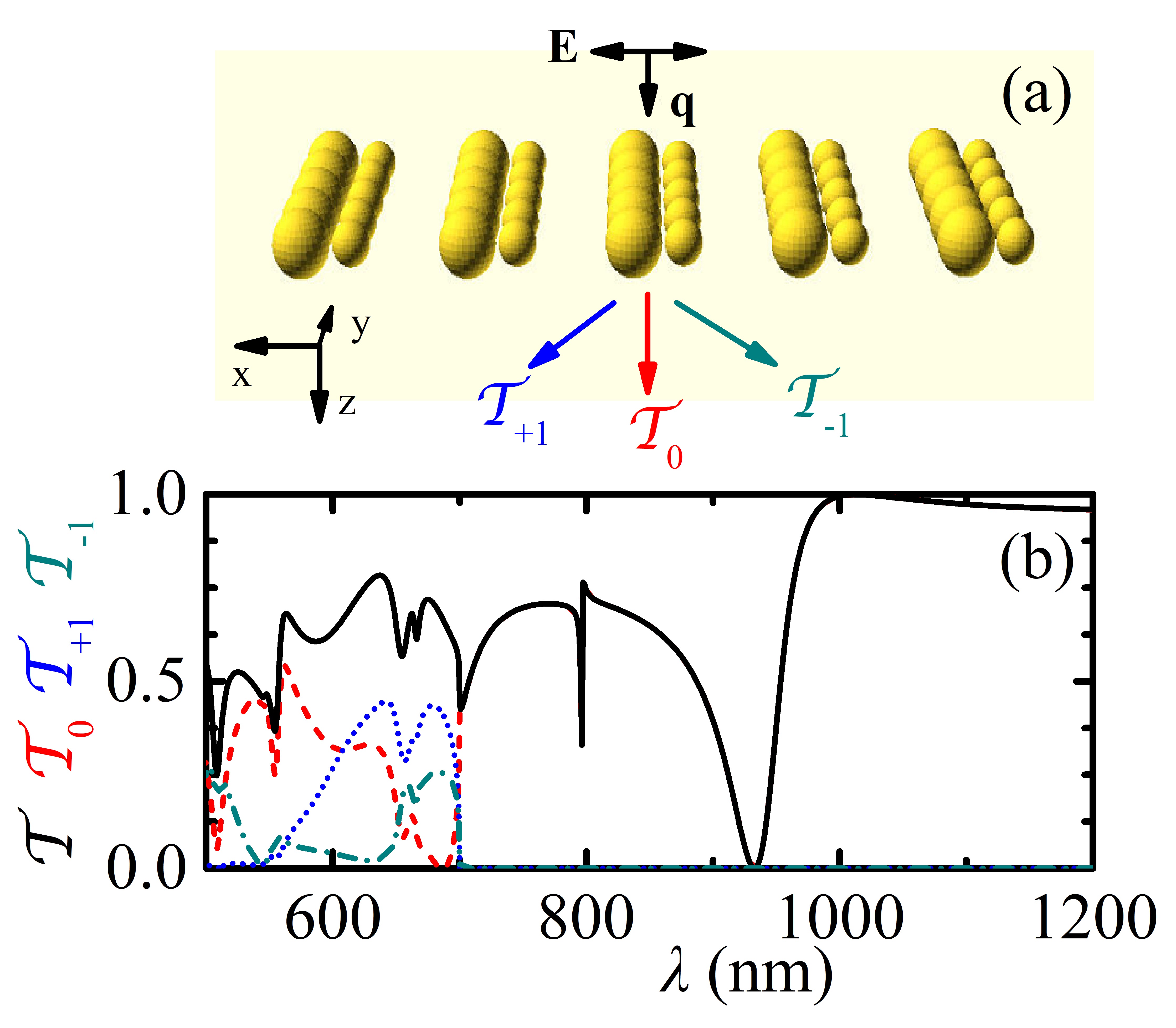}}
\caption{(a): An array of Si sphere dimers with radii $r_1=120$~nm and $r_2=80$~nm, with their centers on the same plane and a separation of 10~nm along the $x$ direction, arranged in a rectangular lattice with $a_x=700$~nm and $a_y=300$~nm. 
(b): Transmittance of the structure described in (a) for normally incident $x$-polarized light. Solid black line: Total transmittance, $\cal{T}$. Dashed red line: Transmittance through zeroth-order diffraction, ${\cal T}_0$ (specular transmittance). Dotted blue line: Transmittance in the $m_x=+1$ diffraction channel,  ${\cal{T}}_{+1}$. Dash-dotted green line: Transmittance in the $m_x=-1$ diffraction channel, ${\cal{T}}_{-1}$.
}
\label{figure006}
\end{figure}

To understand the different underlying mechanisms, we replace the single Si spheres in the rectangular lattice considered previously (see Fig.~\ref{figure003}) with the asymmetric Si sphere dimer $r_1=120$~nm, $r_2=80$~nm, discussed above. The geometry is schematically shown in Fig.~\ref{figure006}(a) and  corresponding transmission spectra for the different diffraction orders, for light incident normally and polarized along the $x$ direction, are depicted in Fig.~\ref{figure006}(b).
These spectra should be compared with those of the lattice with 
the single spheres, for the same polarization, depicted in Fig.~\ref{figure003}(a). It can be seen that the sharp drop in the transmission just above $900$~nm due to the DM mode of the larger spheres is preserved also
in the dimer lattice. The Mie resonances of the smaller spheres affect the spectra below $700$~nm, but the characteristic drop of the specular transmittance, ${\cal T}_0$, just below 
700~nm, attributed to the DE mode of the larger sphere, appears in the spectra of both geometries. Additionally, the first-order diffracted beams in the $x$ direction 
($m_x=\pm 1$) 
now become asymmetric. More power is directed in the ${\cal T}_{+1}$ beam and 
 light is bent in the direction of the larger spheres, which
is consistent with the scattering pattern of the isolated 
pair [see Fig.~\ref{figure005}(b)]. Higher-order diffraction 
($|m_x|>1$) is negligible in the spectral window considered. 
A strongly localized leaky mode just below $\lambda=800$~nm, which yields a sharp transmission dip, is also worth-mentioning. The existence of such Fano-type modes has been discussed in the literature~\cite{fano1,fano2}, and their position and strength depends on their interaction with the Mie modes of the spheres.

The most interesting spectral region 
for our purposes appears below the diffraction edge, at $\lambda$ ranging from 650~nm to 700~nm, where the interaction of the DE Mie modes of the larger spheres strongly suppresses the specular transmission but, at the same time, the spectral overlap of the DE and DM modes of the different spheres (Kerker's condition) minimizes the reflectance. As a result, most of the power is transmitted through the first-order diffraction channels. 
The asymmetry 
between the first-order diffraction channels in the $x$ direction,
${\cal T}_{+1}$ and ${\cal T}_{-1}$, is not large but can be further optimized 
given the far-field scattering amplitude of the isolated Si pair [see Fig. \ref{figure005}(b)].

\begin{figure}[htbp]
\centering
\fbox{\includegraphics[width=0.5\linewidth]{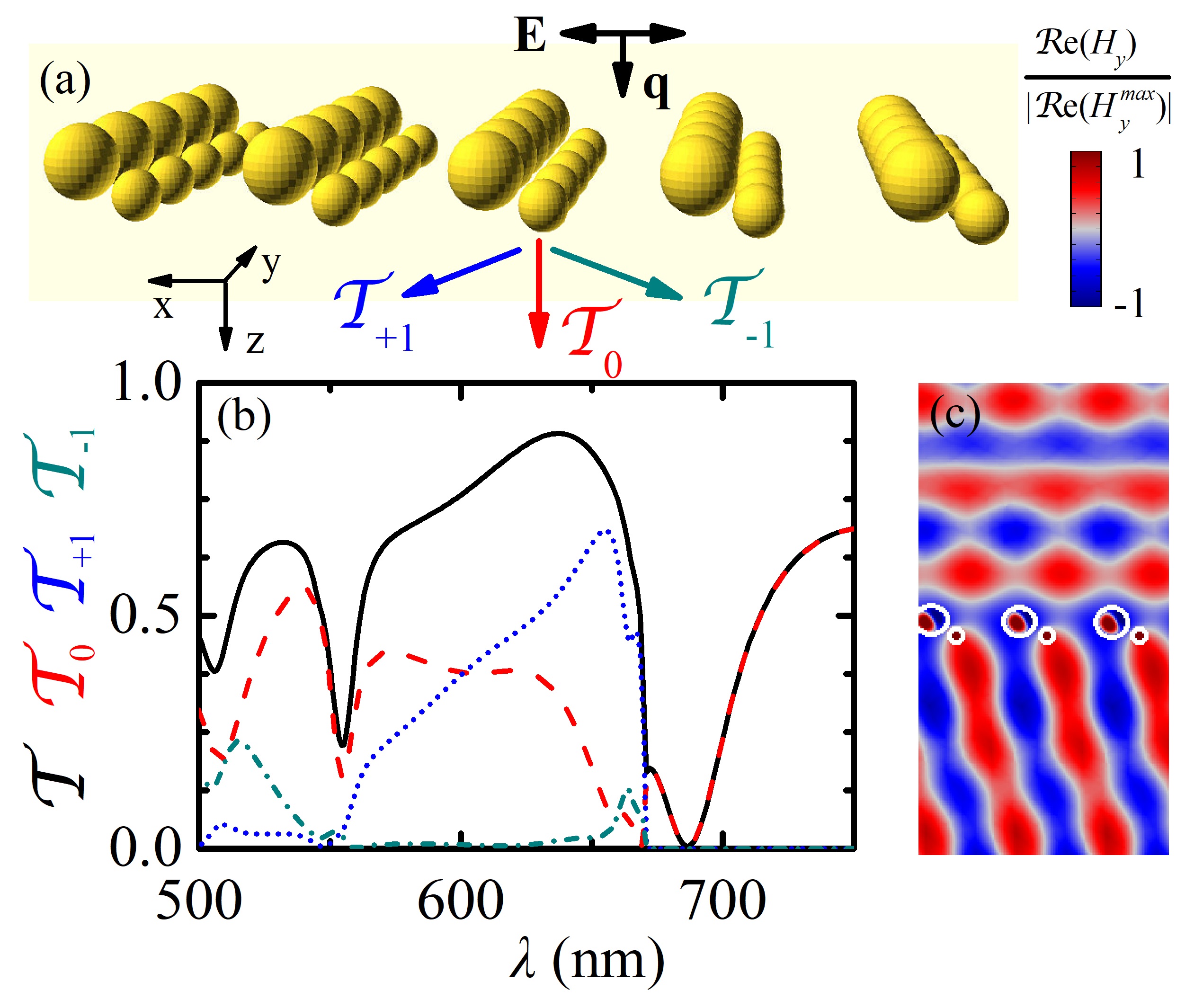}}
\caption{ (a): A rectangular array, with $a_x=670$~nm and $a_y=270$~nm, of the Si sphere dimers of Fig.~\ref{figure006}(a) rotated 
by an angle $\theta=28^o$ about an axis parallel to the $y$ direction passing through the centers of the larger spheres.
(b): Transmittance of the structure described in (a) for normally incident $x$-polarized light. Solid black line: Total transmittance, $\cal{T}$. Dashed red line: Transmittance through zeroth-order diffraction, ${\cal T}_0$ (specular transmittance). Dotted blue line: Transmittance in the $m_x=+1$ diffraction channel,  ${\cal{T}}_{+1}$. Dash-dotted green line: Transmittance in the $m_x=-1$ diffraction channel, ${\cal{T}}_{-1}$. (c) Profile of the $y$ component of the magnetic field, $H_y$, in an $x$-$z$ plane passing through the centers of the spheres, at $\lambda=650$~nm. }
\label{figure007}
\end{figure}

Let us consider a rectangular array, with lattice parameters $a_x=670$~nm and $a_y=270$~nm, with the Si dimers rotated 
by an angle $\theta=28^o$ about an axis parallel to the $y$ direction passing through the centre of the larger spheres, as shown in Fig.~\ref{figure007}(a). This structure is optimized for maximum intensity of directional transmission and larger deflection angle, through extensive numerical calculations by the LMS method.
The relevant transmission spectra associated with the different diffracted beams are displayed in Fig.~\ref{figure007}(b).  
Transmission towards the larger sphere is maximized while leakage of light in all other channels is minimized at wavelengths about 660~nm and ${\cal T}_{+1}$ reaches a peak close to 70\% while the bending angle at $\lambda=650$~nm is $\varphi_{+1}=76^\mathrm{o}$. The calculated magnetic-field profile at this wavelength is shown in Fig.~\ref{figure007}(c).
Interestingly, strong diffraction persists with more than 50\% of the incoming power funneled in the $m_x=+1$ channel in a relatively broad spectral window ranging from about 630~nm to 670~nm.

\begin{figure}[htbp]
\centering
\fbox{\includegraphics[width=0.5\linewidth]{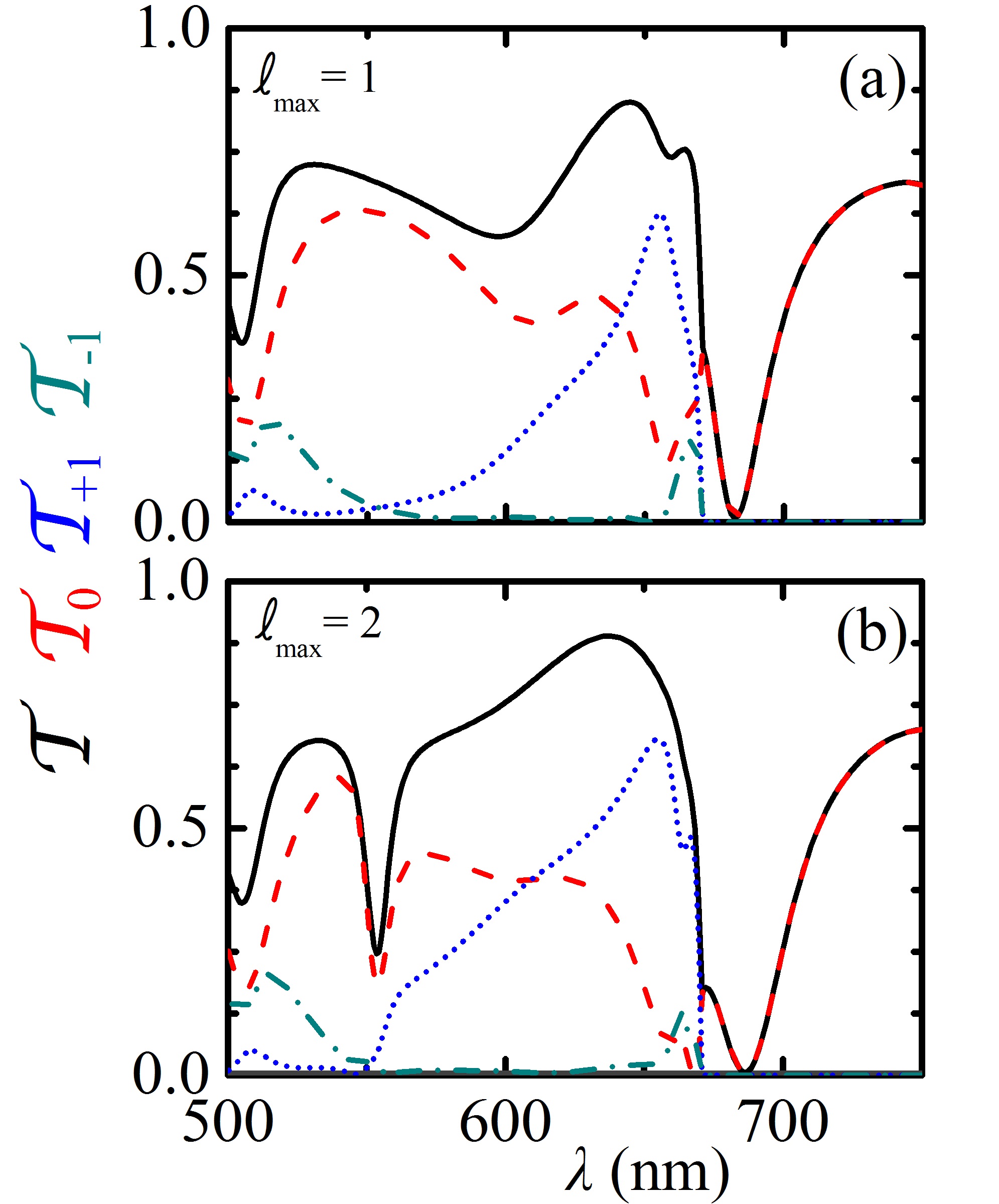}}
\caption{The spectra of Fig.~\ref{figure007}(b), calculated by taking into account (a): Only dipolar terms (${\ell}_{\mathrm{max}}=1$). (b): Dipolar and quadrupolar terms (${\ell}_{\mathrm{max}}=2$). }
\label{figure008}
\end{figure}

 A comment on the necessity of a full multipole description of the effects discussed here above is relevant.
 As mentioned previously, all higher multipoles are included in our calculations and the results reported in the present work are fully converged. On the other hand, dipole models are often used and assumed sufficient~\cite{Si_dimers_1} to describe the optical response of isolated Si dimers of similar size noting that quadrupole modes, though they result in strongly anisotropic scattering, have longer lifetimes and thus exhibit higher losses compared to the dipole modes. 
The influence of the different multipoles on the results for the structures studied here is shown in Fig.~\ref{figure008}, where we present spectra calculated
for our optimized Si sphere dimer structure of Fig.~\ref{figure007}(a) by truncating the multipole expansions involved at $\ell_{\mathrm{max}}=1$ and $\ell_{\mathrm{max}}=2$. It can be seen that, qualitatively, the overall behavior of the system can be described by the dipole approximation
since both the suppression of the specular transmission and the strong asymmetry of the first-order diffraction channels are reproduced. 
However, quantitative information requires to include quadrupole ($\ell=2$) terms while $\ell>2$ give only small contributions and can be ignored.

 Our structures achieve strong and efficient light bending 
 and compare very well to other metasurfaces proposed for this purpose, as can be seen in Table~\ref{table1}, where $\mathcal{T}_{\mathrm{max}}$ and $\eta$ are the maximum transmittance and light-deflection angle obtained.
 The present design is simple, optimization was based on 
 physical insight, and similar principles can be used to 
 design efficient deflectors at different wavelengths by appropriately adjusting the material and geometrical parameters involved.      
Although we used air as embedding medium for our proof of principle design, following the same optimization procedure, similar efficiencies can be obtained in glass, at somewhat longer wavelengths in the near infrared.

%\subsection{Sample Table}

%Table \ref{tab:shape-functions} shows an example table.

\begin{table}[htbp]
\centering
\caption{Comparative performance of different designs of transmissive metasurfaces for large-angle unidirectional deflection of light.}
\begin{tabular}{ccccc}
\hline
Ref. & $\eta$ & $\cal{T}_{\mathrm{max}}$ & Frequency & Complexity\\
\hline
\cite{metagrat_diel_bianisotropic} & $70^{\mathrm{o}}$ & $95 \%$ & mid-infra & multi-parametric\\
\cite{freeform} & $75^{\mathrm{o}}$ & $84 \%$ & near-infra & extremely complex\\
\cite{fishlike} & $80^{\mathrm{o}}$ & $50 \%$ & visible & very complex\\
\cite{fishlike} & $55^{\mathrm{o}}$ & $90 \%$ & visible & very complex\\
\cite{metalens} & $82^{\mathrm{o}}$ & $50 \%$ & visible & Si-cylinder dimer  \\
This work & $76^{\mathrm{o}}$ & $70 \%$ & visible & Si-sphere dimer in air\\
\hline
\end{tabular}
  \label{table1}
\end{table}

\section{Conclusion}

In summary, we developed a strategy to engineer metasurfaces capable of strongly and efficiently deflecting polarized transmitted light. This consists in first designing a periodic array of high-refractive-index nanoparticles that block the forward transmission channel and, subsequently, add a second smaller particle per unit cell in order to guide light asymmetrically through constructive interference to the desired diffraction order. Our approach takes advantage of the simultaneous existence of DM and DE resonant modes in such particles, and the deflection angle 
can be maximized by adjusting the geometric parameters involved to tune interference effects 
close to the diffraction edge of the lattice. The efficiency of this strategy was demonstrated on simple periodic arrays of asymmetric Si sphere dimers where, by means of systematic numerical calculations using an extension of the LMS method in conjunction with finite-element simulations, we provided unambiguous evidence for large-angle unidirectional deflection of visible light. The underlying design principles, which are based on the interaction between DM and DE Mie resonances, can be applied to other similar systems such as cylinders on a substrate and establish simple rules for facile optimization of surfaces of arrays of nanoparticles based on the scattering properties of the individual building units.
%\section{Supplemental Material}

%Consult the Author Guidelines for Supplementary Materials in OSA Journals for details on accepted types of materials and instructions on how to cite them.
%All materials must be associated with a figure, table, or equation or be referenced in the results section of the manuscript.
%(1) 2D and 3D image files and video must be labeled “Visualization,” not “Movie,” “Video,” “Figure,” etc.
%(2) Machine-readable data (for example, csv files) must be labeled  “Data File.”  Number data files and visualizations consecutively, e.g., “Visualization 1, Visualization 2….”
%(3) Large datasets or code files must be placed in an open, archival database.  Such items should be mentioned in the text as either “Dataset” or “Code,” as appropriate, and also be cited in the references list.  For example, “see Dataset 1 (Ref. [1]) and Code 1 (Ref [2]).” Here are examples of the references:

%\subsection{Sample Dataset Citation}

%1. M. Partridge, "Spectra evolution during coating," figshare (2014) [retrieved 13 May 2015], http://dx.doi.org/10.6084/m9.figshare.1004612.

%\subsection{Sample Code Citation}

%2. C. Rivers, "Epipy: Python tools for epidemiology" (Figshare, 2014) [retrieved 13 May 2015], http://dx.doi.org/10.6084/m9.figshare.1005064.
                
\section{Funding Information}

We acknowledge support of this work by the project “Development of Materials and Devices for Industrial, Health, Environmental and Cultural Applications” (MIS 5002567) which is implemented under the “Action for the Strategic Development on the Research and Technological Sector”, funded by the Operational Program "Competitiveness, Entrepreneurship and Innovation" (NSRF 2014-2020) and co-financed by Greece and the European Union (European Regional Development Fund).
E.~A. was supported by the General Secretariat for Research and Technology and the
Hellenic Foundation for Research and Innovation under Grant 1819.
E.~P. acknowledges support from IKY PhD scholarship.

%\section{References}

%Note that \emph{Optics Letters} uses an abbreviated reference style. Citations to journal articles should omit the article title and final page number; this abbreviated reference style is produced automatically when the \emph{Optics Letters} journal option is selected in the template, if you are using a .bib file for your references.

%However, full references (to aid the editor and reviewers) must be included as well on a fifth informational page that will not count against page length; again this will be produced automatically if you are using a .bib file.

%\bigskip
%\noindent Add citations manually or use BibTeX. See \cite{Zhang:14,OSA,FORSTER2007,testthesis}.

% Bibliography
%\bibliography{sample}

\begin{thebibliography}{1}
\bibitem{Si_sphere_Mie}For a Si sphere in air the dipolar magnetic Mie
mode is found at $\lambda_{DM} \approx$ 900~nm and the dipolar
electric at $\lambda_{DE} \approx$ 700~nm, while the magnetic quadrupole resonance is
close to $\lambda_{QM} \approx $660~nm.
\end{thebibliography}

% Full bibliography added automatically for Optics Letters submissions; the following line will simply be ignored if submitting to other journals.
% Note that this extra page will not count against page length
%\bibliographyfullrefs{sample}

%Manual citation list

\end{document}